%% file: 18_ieee_lidar.tex
\documentclass[12pt,journal,draftcls,onecolumn]{IEEEtran}

\input{usepackages}
\usepackage{multirow}
\input{definitions}

\newcommand{\mytitle}{LIDAR Data for Deep Learning-Based\\mmWave Beam-Selection}

\begin{document}

\title{\mytitle}

\iftrue


\author{Aldebaro Klautau, Nuria Gonz\'alez-Prelcic and Robert W. Heath Jr.
\thanks{This material is based upon work supported in part by the National Science Foundation under Grant No. ECCS-1711702 and Grant No. CNS-1731658, as well as gifts from Nokia Bell Labs and Toyota ITC. The work of A. Klautau was supported in part by CNPq, Brazil (201493/2017-9/PDE).}
\thanks{A. Klautau (aldebaro@ufpa.br) is with UFPA, Belem, PA, 66075, Brazil.}
\thanks{N. Gonz\'alez-Prelcic (ngprelcic@utexas.edu) and R. Heath (rheath@utexas.edu) are with the Wireless Networking and Communications Group, The University of Texas at Austin, Austin, TX 78712, USA.}
}
\else
\author{Author1, 
				Author2, 
				Author3, 

				Author4, 
				Author5, 
				Author6 
				and Author7
\thanks{XXXXX XXXXX XXXXX XXXXX XXXXX XXXXX XXXXX XXXXX XXXXX XXXXX XXXXX XXXXX 
XXXXX XXXXX XXXXX XXXXX XXXXX XXXXX XXXXX XXXXX XXXXX XXXXX XXXXX XXXXX XXXXX 
XXXXX XXXXX XXXXX XXXXX XXXXX XXXXX XXXXX  
}
\thanks{Author1, Author2 and Author3 are with XXXXXXXXXXXXXXXXXX , XXXXX, XXXXXX (e-mail: \{Author1, Author2,Author3\}@xxxx.xx).}
\thanks{Author4, Author5 and Author6 are with XXXXXXXXXXXXXXXXXX , XXXXX, XXXXXX (e-mail: \{Author4, Author5,Author6\}@xxxx.xx).}
\thanks{Author7 is with XXXXXXXXXXXXXXXXXX , XXXXX, XXXXXX (e-mail: \{author7\}@xxxx.xx).}
}
\fi

\maketitle

\begin{abstract}
Millimeter wave communication systems can leverage information from sensors to
reduce the overhead associated with link configuration.
LIDAR (light detection and ranging) is one sensor widely used in autonomous driving for high resolution mapping and positioning.
This paper shows
how LIDAR data can be used 
for line-of-sight detection
and to reduce the overhead in millimeter wave beam-selection. In the proposed distributed
architecture, the base station broadcasts its position. The connected vehicle
leverages its LIDAR data to suggest a set of beams selected
via a deep convolutional neural network.
Co-simulation of communications and LIDAR in a vehicle-to-infrastructure (V2I) scenario
confirm that LIDAR can help configuring mmWave V2I links.
\end{abstract}

\renewcommand\IEEEkeywordsname{Keywords}

\section{Introduction}

Millimeter wave (\emph{mmWave}) is a key technology for sharing high rate sensor data for 
connected and automated vehicles~\cite{Prelcic17}. Prior work has shown that position information
obtained from vehicles can be used to reduce the overhead required to establish
mmWave links~\cite{capone_obstacle_2015,abbas_context_2016,aviles_position-aided_2016,va_beam_2016-3,Prelcic17,loch_mm-wave_2017,va_inverse_2018}. In this paper, we show how LIDAR provides an additional source
of information to reduce communication overhead. The LIDAR uses a laser to scan an area and measure the time delay
from the backscattered signal. This data is then converted into
points in space and interpreted as three-dimensional (3D) images with
pixels indicating relative positions from the sensor~\cite{choi_millimeter-wave_2016}.
 LIDAR is used on automated vehicles for mapping, positioning, and obstacle detection. 


Reducing the beam-selection overhead is important in cellular and WiFi systems operating at mmWave frequencies~\cite{kim_fast_2014,choi_millimeter-wave_2016,zhou_enhanced_2017}.
Out-of-band measurements were used for improved beam-selection in mmWave communications in \cite{nitsche_steering_2015,ali_millimeter_2017}.
The benefit of a radar located in infrastructure was investigated in \cite{gonzalez-prelcic_radar_2016}.
The use of 
position information in V2I mmWave was studied in~\cite{capone_obstacle_2015,abbas_context_2016,aviles_position-aided_2016,va_beam_2016-3,Prelcic17,loch_mm-wave_2017,va_inverse_2018}.
Some work using position, targeted only line-of-sight (LOS) situations~\cite{va_beam_2016-3,loch_mm-wave_2017,abbas_context_2016}.
Non-LOS (NLOS) was investigated in \cite{aviles_position-aided_2016,va_inverse_2018} with measurement \emph{fingerprint} databases.
Prior work has established that position information can reduce mmWave beam-selection overheads,
and that machine learning (ML) is a good tool for this problem.
But the performance of previously proposed systems is limited by the \emph{penetration rate}  of connected vehicles. The use of LIDAR, which is popular for automated cars, has not been considered, nor have decentralized architectures for applying ML to beam-selection problems.






In this paper, we develop a \emph{distributed} architecture for
reducing mmWave beam-selection overhead. We assume the BS broadcasts
its position via a low-frequency control channel (CC), and
all processing is performed by the connected vehicle. The
vehicle uses its LIDAR data, its own position, and the broadcasted
BS position, to estimate a set of $M$ candidate beam pairs
that are informed to the BS through the CC. The recommended beam pairs are then trained by the BS, and the best one is chosen for data transmission.
 Our system uses only the LIDAR and position information for the prediction; fusion with other sensors is a topic of future work. 

We use ML to solve two key problems in our LIDAR-aided mmWave system. First, we develop a predictor to assess whether the channel is in LOS or NLOS. 
LOS detection is useful because beam-selection is easier in the LOS setting. Second, we use deep learning (DL)~\cite{Lecun2015} with a neural
network trained to perform top-$M$ classification~\cite{geron2017hands} conditioned on LOS and NLOS state estimates. We take this approach instead of alternatives such as \emph{subset ranking}~\cite{Cossock08} because all $M$ selected beams are evaluated in the subsequent stage and, consequently,
their local rank is irrelevant.

We present simulation results obtained with a methodology that combines a traffic simulator to model realistic mobility
scenarios with integrated (``paired'') data from ray-tracing (for
estimating mmWave channels) and LIDAR simulators. Our
results indicate that the beam-selection overhead can be reduced
by factors of 12x in LOS and 2x in NLOS, without
reduction of throughput or by larger factors if some reduction
is acceptable. Compared with prior work~\cite{capone_obstacle_2015,abbas_context_2016,aviles_position-aided_2016,va_beam_2016-3,Prelcic17,loch_mm-wave_2017,va_inverse_2018}, we consider LIDAR on the vehicle as an additional sensor. We also use DL because of its promising
results for position-based beam-selection~\cite{klautau_5g_2018} and many other domains~\cite{Lecun2015,geron2017hands}. An advantage of our approach versus~\cite{capone_obstacle_2015,abbas_context_2016,aviles_position-aided_2016,va_beam_2016-3,Prelcic17,loch_mm-wave_2017,va_inverse_2018} is that our distributed architecture does not depend on the penetration rate of connected vehicles, as it only uses the LIDAR of the connecting vehicle.

\section{System Model}

%





We consider a downlink OFDM mmWave system with analog beamforming~\cite{ali_millimeter_2017}. Both transmitter and receiver have antenna arrays with only one radio frequency (RF) chain and fixed beam codebooks. 
To simulate the channel, we use ray-tracing data and combine the ray-tracing output with a \emph{wideband}  mmWave \emph{geometric channel model} as, e.\,g., in~\cite{ali_millimeter_2017}.
Assuming 
$R_c$ multipath components (MPC) per transmitter / receiver pair, the information collected from the  outputs for the $r$-th MPC of a given pair is: complex path gain $\alpha_{r}$, time delay $\tau_{r}$ and angles $\phi_r^D$, $\theta_r^D$, $\phi_r^A$, $\theta_r^A$, corresponding respectively to azimuth and elevation for departure and arrival.
The frequency-selective channel model at the time instant corresponding to the $n$-th symbol vector is described in detail in~[Section III]\cite{ali_millimeter_2017},  which also includes the definition of the model in the frequency domain $\mathbf{H}[k]$, where $k$ is the subcarrier index.

\begin{align}
\mathbf{H}[n] = \sqrt{N_t N_r}\sum_{\ell = 0}^{L_p-1} \alpha_{\ell} g(nT_s-\tau_{\ell}) \mathbf{a}_r(\phi_{\ell}^A, \theta_{\ell}^A)\mathbf{a}^*_t(\phi_{\ell}^D, \theta_{\ell}^D), 
\end{align}
where $N_t$ and $N_r$ are the numbers of antennas at the transmitter and receiver, respectively, $g(\tau)$ is the shaping pulse (a raised cosine with roll-off of 0.1), $T=1/B$ is the symbol period and $B$ the bandwidth, $\mathbf{a}_r(\phi_r^A, \theta_r^A)$ and $\mathbf{a}^*_t(\phi_r^D, \theta_r^D) $ are the steering vectors at the receiver and transmitter for the $r$-th MPC, respectively. 
Assuming OFDM with $K$ subcarriers and that $\mathbf{H}[n]$ can be accurately represented by its first $L$ taps, the frequency-domain channel at subcarrier $k$ is
\begin{align}
\mathbf{H}[k] = \sum_{n = 0}^{M-1} \mathbf{H}[n] e^{-j\frac{2 \pi k}{K}n}.
\end{align}
\begin{align}
\bH_v = \bU_r^* \textrm{~} \bH \textrm{~} \bU_t
\end{align}

$$
K < N / \log_2(N)
$$

$$
\textrm{SQNR} = -10 \log_{10} \textrm{NMSE}
$$

We assume beam codebooks $\mathcal{C}_t = \{\mathbf{f}_1, \cdots, \mathbf{f}_{|\mathcal{C}_t|}\}$ and $\mathcal{C}_r= \{\mathbf{w}_1, \cdots, \mathbf{w}_{|\mathcal{C}_r|}\}$ at the transmitter and the receiver sides,
with no restriction on the codebook size (e.\,g., they do not have to be DFT codebooks). 
For a given pair $(p,q)$ of vectors, representing precoder $\mathbf{f}_p$ and combiner $\mathbf{w}_q$, 
the received signal at subcarrier $k$ is $\bs[k]=\mathbf{w}_q^H \mathbf{H}[k] \mathbf{f}_p$, where $H$ denotes conjugate transpose.
The beam-selection is guided by the normalized signal power
\begin{align}
y_{(p,q)}= \sum_{k = 0}^{K-1} |\mathbf{w}_q^H \mathbf{H}[k] \mathbf{f}_p|^2
\label{eq:beamOutput}
\end{align}
and the \emph{optimum} beam pair is
$
\widehat{(p,q)} = \arg\max_{(p,q)} y_{(p,q)}
$. In this paper, the goal of beam-selection is to recommend a set $\calB = \{(p_i,q_i)\}_{i=1}^M$ such that $\widehat{(p,q)} \in \calB$.

%
%
%

\section{Machine Learning using LIDAR Data}
\label{sec:input_features}


\subsection{Information exchange protocol}

\begin{figure}[htb]
\centering
\includegraphics[width=7cm]{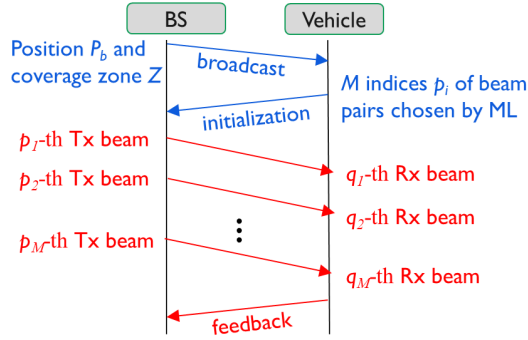}
\caption{Timing diagram for the distributed LIDAR-based beam-selection method. The first phase (broadcast and initialization) uses a low-frequency CC while the second corresponds  to mmWave communication.\label{fig:protocol}}
\end{figure}

We develop a ML-based \emph{beam-selection} strategy for V2I mmWave cellular communication system, assuming that the connected vehicle is equipped with a LIDAR. 
The proposed ML-based protocol is illustrated in \figl{protocol}.
It is assumed that the BS 
can broadcast its absolute position $P_b=(x_b,y_b,z_b)$ for
mmWave V2I beam alignment of incoming vehicles using a CC provided by, for instance, DSRC signals or as part
of the BS CC~\cite{choi_millimeter-wave_2016}. A vehicle estimates its position $P_v = (x_v,y_v,z_v)$
using for example, \emph{Global Positioning System} (GPS) or a simultaneous localization and mapping (SLAM) algorithm~\cite{Narula18}. 
To enable fixed-resolution grids,
the BS also broadcasts its \emph{coverage zone} $Z=(x_1,y_1,x_2,y_2,h)$, which is
the 3D region covered by the BS. The zone $Z$ is
 a cuboid specified by its height $h$, and points $(x_1,y_1)$ and $(x_2,y_2)$ denoting the cuboid base.

The ML algorithm is executed at the vehicle and outputs a set $\calB = \{(p_i,q_i)\}_{i=1}^M$ of beam pairs, where $p_i$ and $q_i$ are indices for precoder and combiner vectors in the predefined codebooks. After this stage,
the $M$ pairs of beams are evaluated at the vehicle, which feedbacks the best one to the BS. 
If \emph{beam correspondence} can be assumed, the same beam pair can be used for uplink.
Once mmWave communication
links are established, the overhead information
required by beam tracking can potentially rely on the high
data rates of mmWave links. 

\subsection{LIDAR-based feature extraction and deep learning}

We use ML to tackle two distinct problems. The first is the use of only LIDAR data for LOS versus NLOS binary classification. The second problem is the selection of the top-$M$ beam pairs based on \equl{beamOutput}
for decreasing the beam-selection overhead, which is associated with the protocol explained in the previous subsection. 
The raw input data to solve both problems
is composed of the LIDAR point cloud $\bC$ collected by the vehicle, the BS coverage zone $Z$ and
positions $P_v$ and $P_b$. The LIDAR cloud $\bC$ is an array of dimension $D \times 3$, composed of 3D points indicating the presence of obstacles.
Typical values of $D$ are relatively large and using an alternative representation helps to control the computational cost.
For example, each point cloud used in this paper originally had at least $D=35,822$ $(x,y,z)$ points.

\begin{figure}[htb]
\centering
\includegraphics[width=8cm]{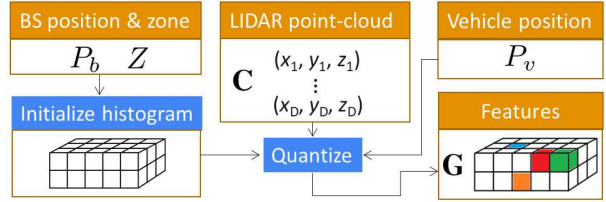}
\caption{Feature extraction of 3D histogram $\bG$ from LIDAR data.\label{fig:feature_extraction}}
\end{figure}


In this paper, we adopt a fixed grid $\bG$ to represent the whole zone $Z$, as depicted in \figl{feature_extraction}. We use $\bG$ as a 3D histogram in which a bin 
corresponds to a fixed region of $Z$. Each element of $\bG$ stores the number of elements of $\bC$ within the corresponding bin. A large count of occurrences indicates that LIDAR detected many points within the bin. Note that the element of $\bG$ corresponding to the position $P_b$ of the BS is the same in all examples given the grid is fixed.
This histogram calculation was implemented as the uniform quantization using $(b_x, b_y, b_z)$ bits
of the elements of $\bC$. Outliers in $\bC$ are discarded in order to design quantizers with adequate dynamic ranges.
We also discard points that are farther from the vehicle (at position $P_v$) by more than a certain distance $d_{\textrm{max}}$.
The ML input feature in then a 3D histogram represented by a sparse matrix $\bG$ of dimension $2^{b_x} \times 2^{b_y} \times 2^{b_z}$.






For both problems (LOS decision and beam-selection), we adopted neural networks with 13 layers from which 7 are 2D \emph{convolutional} layers with
decreasing kernel sizes, from $13 \times 13$ to $3 \times 3$, trained
with 
Kera's \emph{Adadelta} optimizer~\cite{geron2017hands}.
We
used \emph{pooling} layers and, to mitigate overfitting, \emph{regularization} and \emph{dropout}.
For beam-selection, the values in \equl{beamOutput} below 6~dB from the maximum
were zeroed and normalized to have unitary sum.
For top-$M$ classification, the output layer had a \emph{softmax} activation function and 
a \emph{categorical cross-entropy} as loss function~\cite{geron2017hands}. For binary classification,
the output layer and loss were 
\emph{sigmoid} and \emph{binary cross-entropy}, respectively~\cite{geron2017hands}.
The number of parameters per network is approximately $10^5$.

As a baseline for comparing with DL applied to the LOS decision problem, we 
also evaluated 
a simple geometric approach: 
given $P_b$ and $P_v$, we calculate the line $\calL$ connecting them.
We denote by $\hat {d}$ the \emph{minimum distance} between any point in $\bC$ to $\calL$.
A \emph{decision stump} classifier~\cite{geron2017hands} uses a threshold $\gamma$ to decide for NLOS if $\hat{d} < \gamma$ or LOS otherwise. 
The intuition is that if $\calL$ is far from all obstacles identified by the LIDAR in $\bC$, the link is potentially LOS. 

\section{Numerical Results}
\label{sec:results}


\subsection{Simulation methodology}
Aiming at realistic datasets, we adopted a simulation methodology using traffic, ray-tracing and LIDAR simulators in V2I mmWave communications~\cite{klautau_5g_2018}.
We paired the simulations of the mmWave communication system and the LIDAR data acquisition integrating three softwares: the \emph{Blender Sensor Simulation} (BlenSor)~\cite{gschwandtner11}, the \emph{Simulation of Urban MObility} (SUMO) traffic simulator~\cite{SUMO2012}, both open source, and Remcom's Wireless InSite for ray-tracing. 
In the configuration stage, the user provides information about the objects in the 3D scenario, lanes coordinates, eletromagnetic parameters, etc. 
The software execution is based on a Python \emph{orchestrator} code that invokes SUMO and converts its ouputs (vehicles positions, orientations, etc.) 
to formats that can be interpreted by distinct simulators. The orchestrator then invokes the simulators (LIDAR and ray-tracing in this case) to obtain paired results.

\begin{figure}[htb]
\centering
\includegraphics[width=8cm]{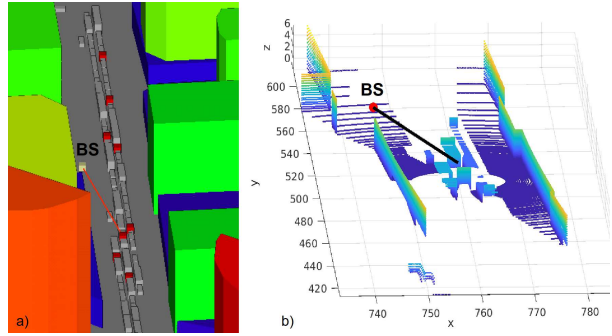}
\caption{a) Urban canyon 3D scenario with 
vehicles of distinct sizes randomly positioned. The building color indicates height and corresponds to a range from 0 (blue) to 101 meters (red). b) Corresponding LIDAR point cloud. The LOS rays between the BS antenna at $z=4$~m and vehicle are shown. \label{fig:rt_and_lidar}}
\end{figure}

Fig.~\ref{fig:rt_and_lidar}a depicts the adopted urban canyon 3D scenario,
which is part of Wireless InSite's examples
and
represents a region of Rosslyn, Virginia.
The \emph{study area} is a rectangle of approximately $337 \times 202 \textrm{~m}^2$
 and the BS antenna array height is $z=4$~m.
We placed receivers and LIDARs on top of all connected vehicles (identified in red) in each scene snapshot.
\figl{rt_and_lidar}b illustrates an example of the corresponding LIDAR \emph{point cloud}. Lines between BS and vehicle are also shown, and suggest a LOS channel.






The ray-tracing simulations used a maximum
of $L=25$ MPCs per transmitter / receiver pair, isotropic antennas, 60~GHz carrier frequency, $B=100$~MHz, $K=64$ subcarriers, and enabled ray-tracing diffuse scattering. Other parameters followed the ones in~\cite{klautau_5g_2018}.

The downlink mmWave massive MIMO relied on a BS with a
$16 \times 16$ uniform planar array (UPA) and vehicles with $4 \times 4$ UPAs. 
When designing $\mathcal{C}_t$ and $\mathcal{C}_r$, we first augmented the conventional DFT codebook with steered codevectors, linear combinations of codevectors and random vectors from Grassmannian codebooks. From this large initial set, we kept only the codevectors that were chosen as $\widehat{(p,q)}$ more than 100 times in the training set. This procedure led to $|\mathcal{C}_t| = 20$ and $|\mathcal{C}_r| = 12$, respectively. Hence, the number of classes for top-$M$ classification is 240.

The LIDAR simulations assumed a Velodyne model HDL-64E2 scanner positioned at a height $z=1$~m from the top-center of the vehicle. 
The angle resolution was 0.1728 degrees and the rotation speed 10~Hz.
The experiments adopted $b_x=b_y=6$ and $b_z=3$ bits.
We eliminated from $\bC$ the points with small values in the $z$-axis ($< 0.1$~m), which correspond to ground
reflections (see \figl{rt_and_lidar}b), and also the points with a distance from the LIDAR larger than $d_{\textrm{max}}=25$~m.

The mmWave channel was assumed noise-free but we considered two conditions with respect to positioning accuracy: \emph{noise-free} and \emph{noisy}.
The LIDAR noise~\cite{gschwandtner11} is assumed to have independent
components distributed according to a zero-mean Gaussian $\calN(0,\sigma^2_L/3)$ with
variance $\sigma^2_L/3$. 
For the \emph{noisy} condition, we adopted the HDL-64E2 default value of $\sigma_L=0.1$~m.
Similarly, the accuracy of the \emph{Global Navigation Satellite System} (GNSS) technology is modeled assuming the elements of the position error vector 
are independent and identically distributed according to $\calN(0,\sigma^2_G/3)$ (no bias).
Conventional GPS may lead to errors of 3 to 5~m, while sophisticated
SLAMs can help to keep the error below 50~cm in the horizontal plane~\cite{Narula18}.
For the \emph{noisy} condition, we assumed $\sigma_G=3$~m and $\sigma_L=0.1$~m.

Beam-selection is harder in NLOS because the predictability decreases considerably when compared to
LOS cases. If an experiment considers both LOS and NLOS channels, the accuracy of ML will depend on the blockage probability,
which is heavily influenced by traffic statistics, 
large vehicles (potential blockers) and antenna height.
Numerical results of distinct experiments that used mixed LOS and NLOS are harder to compare and the ML models may be biased by the easier LOS cases. To avoid this situation,
we present separate evaluations of beam-selection for each case.
The mmWave data is composed of $N_L=6,482$ LOS and $N_N=4,712$  NLOS channel examples.
The beam-selection experiments used  
$N_L$ and $N_N$ examples in the LOS and NLOS evaluations, respectively, while 
LOS detection used all $N_L+N_N$ examples.
For all experiments we created disjoint test and training sets with 20\% and 80\% of the examples, respectively.

\subsection{Results}

The \emph{accuracy} of both binary and top-$M$ classifiers improve
considerably 
when the elevation angle of the LIDAR 
is adjusted for communications (points to the BS antenna). We did not perform this adjustment 
and used the HDL-64E2 default elevation. 
%
%
This increases 
the chances that the LIDAR does not detect a LOS blocker 
because it is obstructed by a neighbor vehicle.
For the LOS detection in noise-free condition, the 
minimum achieved misclassification error with the geometry-based decision stump was 24\% while DL leads to 10\%.

\begin{figure}[htb]
\centering
\includegraphics[width=6.5cm]{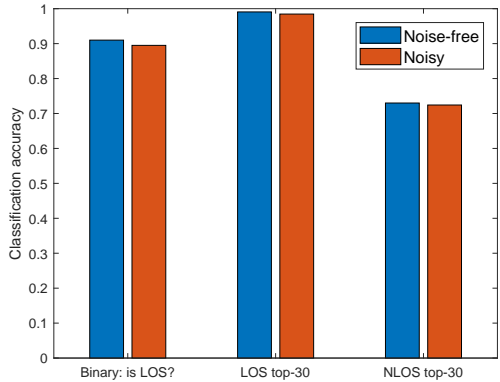}
\caption{Accuracy for LOS detection (binary problem) and beam-selection using top-$M$ classification with $M=30$ for LOS and NLOS examples. The performance for both noise-free and noisy conditions are shown.\label{fig:noise_impact}}
\end{figure}

\figl{noise_impact} presents the results using DL for LOS detection and the two cases of top-$M$ beam-selection for both
(positioning) noise scenarios. It can be seen that the adopted noisy condition did not lead to significant loss of accuracy.
As expected, the performance in NLOS is considerably lower than for LOS.
Due to the difficulty of dealing with NLOS, the binary problem has worse performance
than top-30 LOS classification.
 
\begin{figure}[htb]
\centering
\includegraphics[width=7.8cm]{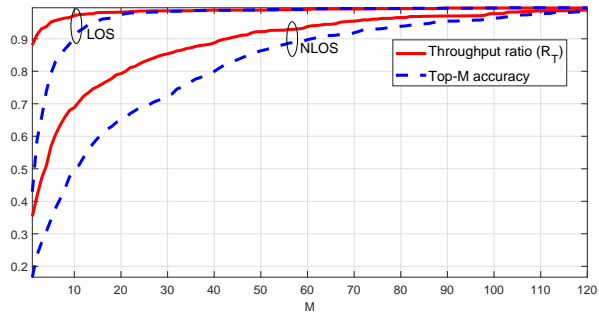}
\caption{LOS and NLOS top-$M$ classification for beam-selection with 240 beam pairs in noisy condition for $M=1,\ldots,120$.\label{fig:acc_rt_shorter}}
\end{figure}

While \figl{noise_impact} shows results for $M=30$ only, \figl{acc_rt_shorter} presents the top-$M$ accuracy for $M=1,\ldots,120$.
 \figl{acc_rt_shorter} also depicts the corresponding \emph{achieved throughput ratio}
\begin{equation}
R_T = \frac{\sum_{i=1}^N \log_2 (1 + y_{\widetilde{(p,q)}})}{\sum_{i=1}^N \log_2 (1 + y_{\widehat{(p,q)}})},
\label{eq:rate_ratio}
\end{equation}
where $N$ is the number of test examples and $\widetilde{(p,q)}$ is the best beam pair in $\calB$. 
For $M=10$, $R_T = 0.97$ and $0.69$ for LOS and NLOS, respectively. In this case, while the 
overhead for beam-selection decreases by a factor of 24, the corresponding $R_T$
indicates a reduction to 69\% of the achievable throughput for NLOS. For NLOS, $R_T$ reaches e.\,g. 94\% for $M=60$.

%

\section{Conclusions}
LIDAR can be used for LOS detection and to reduce the mmWave beam-selection overhead in V2I scenarios.
The results are promising in spite of the relatively simple adopted features.
Future work includes exploring alternative features, fusing data from LIDAR and others sensors,
using a larger amount of data, and better 
tuning the many ML parameters for improved NLOS performance.

%
%
%
%
%
%
%
%
%
%

\bibliographystyle{IEEEtran}
\bibliography{IEEEabrv,references,zotero_exported_items_processed} 

\end{document}

%% file: usepackages.tex
\ifdefined\usePortuguese
\usepackage[brazil]{babel}   
\usepackage[latin1]{inputenc}
\else
\fi

\ifdefined\appendices
\else
\ifdefined\addAppendixAsPreamble 
\usepackage[titletoc,title]{appendix} 
\else
\usepackage{appendix} 
\fi
\fi

\usepackage{makeidx}  
\usepackage{graphicx}
\usepackage{color} 
\definecolor{darkgreen}{rgb}{0,0.35,0}

\ifdefined\keywords
\else

	\def\keywords{\vspace{.5em}{\bfseries\textit{Index Terms}---\,\relax%
	}}
	
\fi

\usepackage{url} 

\ifdefined\RCSfile 
\else
\usepackage{cite} 
\fi

\usepackage{multirow} 

\usepackage[bbgreekl]{mathbbol}

\ifdefined\amsmath
\else
\usepackage{amsmath}
\fi

\ifdefined\amsfonts
\else
\usepackage{amsfonts}
\fi

\usepackage{amssymb}

\ifdefined\proof
\else
\usepackage{amsthm}
\fi

\ifdefined
\else
\ifdefined\akbook
\ifdefined\lulu
\typearea[6mm]{1} 
\else
\usepackage[a4paper, top=3cm, bottom=3.4cm, left=1.8cm, right=2.5cm]{geometry}
\fi
\fi
\fi

\usepackage{mathrsfs} 
\usepackage{makeidx}  
\usepackage{graphicx,color}

\usepackage{multirow} 
\usepackage{listings} 
\usepackage{ifpdf} 

\ifpdf

\ifdefined\hyperref
\else
\usepackage[pdftex,colorlinks]{hyperref}
\fi

	\usepackage[pdftex]{thumbpdf} 
	\usepackage{pdflscape}
\usepackage{epstopdf}
\hypersetup{%
pdftitle={Some title},
pdfauthor={Your name - LaPS - UFPA},
pdfkeywords={DSP,Signal},
pdfstartview={FitH}, 
urlcolor=black,
linkcolor=black,
citecolor=black,
}

\fi 



%% file: definitions.tex
\newcommand{\equl}[1] {{(\ref{eq:#1})}}
\newcommand{\figl}[1] {{Fig.~\ref{fig:#1}}}

\newcommand{\bH}{{\bf H}}

\ifdefined\bm
\else
\newcommand{\bm}[1]{{\mbox{\boldmath $#1$}}}
\fi

\ifdefined\Theorem
\else
\newtheorem{Theorem}{Theorem}
\fi

\newtheorem{Theorem*}{Theorem}

\newtheorem{Application*}{Application}

\newtheorem{Claim*}{Claim}

\ifdefined\Corollary
\else
\newtheorem{Corollary}[Theorem]{Corollary}
\fi

\newtheorem{CounterExample*}{$\overline{\hbox{\bf Example}}$}

\ifdefined\Example
\else
\newtheorem{Example}{Example}
\fi

\newtheorem{Example*}{Example}

\newtheorem{Intuition*}{Intuition}
\newtheorem{Joke*}{Joke}

\ifdefined\Lemma
\else
\newtheorem{Lemma}[Theorem]{Lemma}
\fi

\newtheorem{Lemma*}{Lemma}

\newtheorem{Model*}{Model}
\newtheorem{Open problem}{Open problem}

\newtheorem{Question*}{Question}
\newtheorem{Remark*}{Remark}

\def \bSubexa    {\begin{subexa}}

\ifdefined\Problem
\else
\newenvironment{Problem}{\textbf{Problem}\\ \begin{enumerate}}{\end{enumerate}}
\fi

\newcommand{\ignore}[1]{{}}

\def \bC {{\textbf C}}

\def \bG {{\textbf G}}
\def \bH {{\textbf H}}

\def \bU {{\textbf U}}

\def \bf {{\textbf f}}

\def \bm {{\textbf m}}

\def \bs {{\textbf s}}

\def \calB {{\cal B}}

\def \calL {{\cal L}}

\def \calN {{\cal N}}

\def \mynote#1{{}}

\def\ignore#1{}

\def\orpro{\mathop{\mathchoice
   {\vee\kern-.49em\raise.7ex\hbox{$\cdot$}\kern.4em}
   {\vee\kern-.45em\raise.63ex\hbox{$\cdot$}\kern.2em}
   {\vee\kern-.4em\raise.3ex\hbox{$\cdot$}\kern.1em}
   {\vee\kern-.35em\raise2.2ex\hbox{$\cdot$}\kern.1em}}\limits}

\def\andpro{\mathop{\mathchoice
 {\wedge\kern-.46em\lower.69ex\hbox{$\cdot$}\kern.3em}
 {\wedge\kern-.46em\lower.58ex\hbox{$\cdot$}\kern.25em}
 {\wedge\kern-.38em\lower.5ex\hbox{$\cdot$}\kern.1em}
 {\wedge\kern-.3em\lower.5ex\hbox{$\cdot$}\kern.1em}}\limits}

